\documentclass[final,12pt]{clear2025} 

\PassOptionsToPackage{round,compress}{natbib}

\usepackage[utf8]{inputenc} 
\usepackage[T1]{fontenc}    
\usepackage{hyperref}%
\usepackage{url}            
\usepackage{booktabs}       
\usepackage{xcolor,colortbl}
\usepackage{graphicx}
\usepackage{multirow}
\usepackage{xfrac}
\usepackage{anyfontsize}
\usepackage[font=small]{caption}
\usepackage{subcaption} 
\usepackage[shortlabels]{enumitem}
\usepackage{comment}
\usepackage{wrapfig}

\hypersetup{colorlinks=true, linkcolor=[rgb]{0.8,0,0}, citecolor=[rgb]{0,0,0.8}, urlcolor=[rgb]{0.8,0,0.8}}
\usepackage{mathtools} 
\usepackage{amsfonts}       
\usepackage{amssymb}
\usepackage{nicefrac}       
\usepackage{microtype}      
\usepackage{physics}
\usepackage[noabbrev,nameinlink]{cleveref}
\usepackage{algorithm, algpseudocode}
\usepackage[skip=3pt plus1pt, indent=20pt]{parskip}

\renewcommand{\labelenumi}{(\roman{enumi})}

\makeatletter
\def\plist@algorithm{Alg.\space}
\makeatother

\usepackage{crossreftools}

\makeatletter
\def\labelText{%
  \@ifstar{\@labeltextstarred}{\@labeltext}
}
\newcommand{\@labeltextstarred}[2]{%
  \crtcrossreflabel*{#1}[#2]%
}

\newcommand{\@labeltext}[2]{%
  \crtcrossreflabel{#1}[#2]%
}
\makeatother

\usepackage{tikz} 
\usetikzlibrary{shapes,decorations,arrows,calc,arrows.meta,fit,positioning,3d,matrix,decorations.pathreplacing}
\tikzset{
    > = stealth,
    every node/.append style = {text = black},
    bidirected/.style={Latex-Latex,dashed},
    directed/.style={-Latex,auto,node distance =1 cm and 1 cm, draw=blue, very thick},
    circled/.style={circle, draw=red, thick},
}

\usepackage{pgfplots}
\pgfplotsset{compat=1.18}
\pgfplotsset{every axis/.append style={
    label style={font=\footnotesize},
    tick label style={font=\footnotesize},
    legend style={font=\footnotesize},
    hide scale/.style={
/pgfplots/xtick scale label code/.code={},
/pgfplots/ytick scale label code/.code={}},
}
}

\usepgfplotslibrary{statistics,fillbetween,colormaps,external,colorbrewer}

\newtheorem{assumption}{Assumption}

\newcommand{\pr}{\mathbb{P}}
\newcommand{\E}{\mathbb{E}}
\newcommand{\R}{\mathbb{R}}

\newcommand{\I}{\mathbb{I}}
\newcommand{\indep}{\perp \!\!\!\! \perp}
\DeclareMathOperator\doo{do}
\DeclareMathOperator\pa{pa}

\let\emptyset\varnothing
\DeclareMathOperator\supp{supp}

\newcommand{\PBc}{\operatorname{PB}_c}
\newcommand{\PHc}{\operatorname{PH}_c}
\newcommand{\PB}{\operatorname{PB}}

\newcommand{\IF}{\mathbb{IF}}

\usepackage{times}

\title[The Probability of Tiered Benefit]{The Probability of Tiered Benefit: Partial Identification\newline with Robust and Stable Inference}

\clearauthor{%
 \Name{Johan {de Aguas}} \Email{johanmd@math.uio.no}\\
 \addr Department of Mathematics, University of Oslo
 \AND
 \Name{Sebastian Krumscheid} \Email{sebastian.krumscheid@kit.edu}\\
 \addr Scientific Computing Center, Karlsruhe Institute of Technology
 \AND
  \Name{Johan Pensar} \Email{johanpen@math.uio.no}\\
 \addr Department of Mathematics, University of Oslo
 \AND
 \Name{Guido Biele} \Email{guido.biele@fhi.no}\\
 \addr Dpt. Child Health \& Development, Norwegian Institute of Public Health
}

\begin{document}

\maketitle

\begin{abstract}%
 We define the \textit{probability of tiered benefit} in scenarios with a binary exposure  and an outcome that is either categorical with $K \geq 2$ ordered tiers or continuous partitioned by $K-1$ fixed thresholds into disjoint intervals. Similarly to other pure counterfactual queries, this parameter is not $g$-identifiable without additional assumptions. We demonstrate that strong monotonicity does not suffice for point identification when $K \geq 3$ and provide sharp bounds both with and without such constraint. Inference and uncertainty quantification for these bounds are challenging tasks due to potential nonregularity induced by ambiguities in the underlying individualized optimization problems. Such ambiguities can arise from immunities or null treatment effects in subpopulations with positive probability, affecting the lower bound estimate and hindering conservative inference. To address these issues, we extend the available \textit{stabilized one-step correction} (S1S) procedure by incorporating stratum-specific stabilizing matrices. Through simulations, we illustrate the benefits of this approach over existing alternatives. We apply our method to estimate bounds on the probabilities of tiered benefit and harm from pharmacological treatment for ADHD upon academic achievement, employing observational data from diagnosed Norwegian schoolchildren. Our findings indicate that while girls and children with low prior test performance could have moderate chances of both benefit and harm from treatment, a clear-cut recommendation remains uncertain across all strata.
\end{abstract}



\section{Introduction}\label{sec:intro}

Counterfactual inference involves probabilistic assessments of events for which certain causal antecedents or consequences are contrary to facts \citep{Balke1994a,Balke1994b}. An exemplar scenario is determining \textit{the probability that a given student would have passed a math qualification test had they taken ADHD medication, given that they did not actually take it and failed the test}. The scope of application of counterfactual queries spans across diverse domains such as epidemiology, economics, legal deliberation, psychology, and AI \citep{PearlCausality}. Notably, it finds high utility in explainability of machine learning \citep{beckers2022causal}, event attribution \citep{climate}, impact assessment \citep{possebom2022}, fairness analysis \citep{Fairness}, and personalized decision-making \citep{MuellerPearl2023}, among other tasks. 

An important counterfactual parameter, in the context of a binary exposure and a binary outcome, is the \textit{probability of necessity and sufficiency} (PNS). This parameter quantifies the proportion of \textit{compliers} in the population, referring to units who would have benefited \textit{if and only if} they had been treated \citep{probcausation}. Typically, this parameter is not $g$-identifiable, meaning it cannot be determined solely from any unconstrained combination of causal graphs, observational data and experiments \citep{Robins89,probcausation}. While point identification may be achievable under a monotonicity assumption \citep{Balke1997}, such a constraint is often unrealistic, as it presupposes the absence of unintended effects from the exposure. Several generalizations of the PNS have been proposed for nonbinary categorical, continuous, and vector-valued exposures and outcomes \citep{Li2024,kawakami2024}. Yet, these have mainly focused on settings with unordered categorical outcomes or on queries that are identifiable under monotonicity.

When $g$-identification is unfeasible, an alternative approach is \textit{partial identification}. Bounds can always be computed from population-level observational and experimental distributions, or solely from the former under the assumption of conditional ignorability \citep{TP}. Yet, conducting inference and uncertainty quantification for estimates of these bounds is challenging due to  potential sources of nonregularity or lack of smoothness in the involved functionals. This difficulty stems from impossibility results indicating that if the target functional is not \textit{pathwise differentiable} at the true distribution, there exist no sequence of \textit{regular} and \textit{locally unbiased} estimators. Furthermore, correction procedures may not fully eliminate bias and could cause the variance to diverge \citep{impossibility}. Lack of regularity and differentiability not only affect the validity of asymptotic inference but also compromise the interpretation of uncertainty quantification under the bootstrap and the Bayesian frameworks \citep{dumbgen1993,fang2019inference,Kitagawa}.

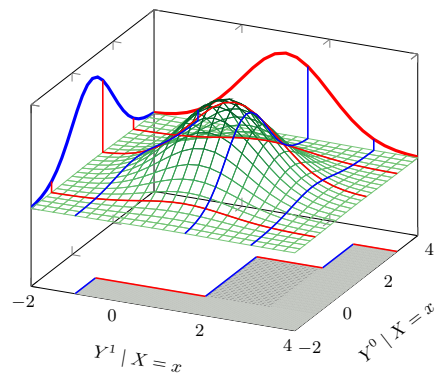
\begin{wrapfigure}{r}{6cm} 
\captionsetup{font=small}
\centering
\begin{tikzpicture}[scale=0.75,
declare function={%
    f(\x,\y)=(1/(2*pi))*exp(-pow(\x-1,2)/2-pow(\y-1,2)/2);}]
 \begin{axis}[
 zmax=0.2, zmin=-0.15,
 colormap/Greens,
 zticklabel=\empty,
 xlabel={$Y^1\mid X=x$},
 ylabel={$Y^0\mid X=x$},
 xlabel style={sloped},
 ylabel style={sloped}]
    \begin{scope}[canvas is xy plane at z=-0.15]
    \draw[color of colormap=900] (1,1) circle[radius=0.5];
    \draw[color of colormap=700] (1,1) circle[radius=1];
    \draw[color of colormap=500] (1,1) circle[radius=1.5];
    \draw[color of colormap=300] (1,1) circle[radius=2];
    \draw[color of colormap=100] (1,1) circle[radius=2.5];
    \end{scope}
    
    \addplot3[surf, samples=30, domain =-1:4, domain y=-2:-1, gray, opacity=0.75]
    {-0.15};
    \addplot3[surf, samples=30, domain =1.5:4, domain y=-1:1.5, gray, opacity=0.75]
    {-0.15};
    \addplot3[surf, samples=30, domain =3:4, domain y=-1:3, gray, opacity=0.75]
    {-0.15};
    
    \draw[color=blue, thick] (-1,-1,-0.15) -- (-1,-2,-0.15);
    \draw[color=blue, thick] (1.5,1.5,-0.15) -- (1.5,-1,-0.15);
    \draw[color=blue, thick] (3,3,-0.15) -- (3,1.5,-0.15);

    \draw[color=red, thick] (-1,-1,-0.15) -- (1.5,-1,-0.15);
    \draw[color=red, thick] (1.5,1.5,-0.15) -- (3,1.5,-0.15);
    \draw[color=red, thick] (3,3,-0.15) -- (4,3,-0.15);
    
    \addplot3[color=red,ultra thick,samples y=0,domain=-2:4] (x,4,{f(x,1)}); 
    \addplot3[color=blue,ultra thick,samples y=0,domain=-2:4] (-2,x,{f(1,x)}); 
    
    \addplot3[mesh,domain=-2:4,domain y=-2:4] {f(x,y)}; 
    
    \addplot3[color=red,thick,samples y=0,domain=4:-2] (x,3,{f(x,3)})
    -- (-2,3,{f(1,3)}); 
    \addplot3[color=red,thick,samples y=0,domain=4:-2] (x,1.5,{f(x,1.5)})
    -- (-2,1.5,{f(1,0.5)}); 
    \addplot3[color=red,thick,samples y=0,domain=4:-2] (x,-1,{f(x,-1)})
    -- (-2,-1,{f(1,-1)}); 
    
    \addplot3[color=blue,thick,samples y=0,domain=-2:4] (-1,x,{f(-1,x)}) --
    (-1,4,{f(-1,1)}); 
    \addplot3[color=blue,thick,samples y=0,domain=-2:4] (1.5,x,{f(1.5,x)}) --
    (1.5,4,{f(1.5,1)}); 
    \addplot3[color=blue,thick,samples y=0,domain=-2:4] (3,x,{f(3,x)}) --
    (3,4,{f(3,1)}); 
 \end{axis}
\end{tikzpicture}

\caption{A setting with a continuous outcome and $K=4, c_1=-1.0, c_2=1.5$, $c_3=3.0$. Here, $\PB(x)$ is the volume under the joint PDF of potential outcomes $(\textcolor{blue}{Y^0},\textcolor{red}{Y^1})| X=x$ enclosed above the benefit region (\textcolor{gray}{gray area}).}
\end{wrapfigure}

\paragraph{Settings:} The studied scenario involves a categorical pre-exposure covariate $X\in\mathcal{X}$ defining strata from the population, a binary point exposure $A\in\{0,1\}$, and an ordered categorical outcome $\tilde{Y}$ with $K\geq 2$ tiers $ C=\{C_k\}_{k=1}^K$. We also consider the case of a continuous outcome $Y\in\mathcal{Y}\subseteq\R$, with support partitioned into $K$ disjoint tiers given by intervals with fixed thresholds $c=\{c_k\}_{k=1}^{K-1}$. We define the \textit{$x$-specific probability of tiered benefit} $\PB(x)$ as the probability that an individual in stratum $x\in\mathcal{X}$ would attain any outcome tier under no treatment and, counterfactually, a higher outcome tier under treatment. Sharp bounds for $\PB(x)$ follow organically from the Fréchet inequalities \citep{frechet1951}. We examine the problem of developing semiparametric estimators for these bounds that are consistent, doubly-robust, and with stable inference in potentially nonregular settings. Our goal is to construct valid \textit{uncertainty regions} for $\PB(x)$, accounting for both systemic and aleatoric uncertainties.

\paragraph{Applied motivation:} Educational and clinical fields are prime domains of application for the \textit{probability of tiered benefit}, as quantitative measurements (e.g. test scores, blood pressure) are often aggregated into ordered tiers or assessed against fixed thresholds for tasks such as preliminary diagnosis, evaluation, categorization, and summarization. For instance, in the Norwegian educational system, compulsory national tests assessing skills in numeracy, reading, and English are administered during school grades 5, 8, and 9. The Norwegian Directorate for Education has established fixed thresholds that group the test scores into \textit{mastery tiers}, ranging from tier 1 (poorest) to tier 5 (highest). Considering the population of Norwegian schoolchildren diagnosed with ADHD and the context of treatment with stimulant medication, one can ask: \textit{What is the probability that a child from a given subpopulation would achieve a higher test mastery tier under treatment than their corresponding counterfactual mastery tier under no treatment?} This is a pure counterfactual query, different from an experimentalist's investigation of the sign and magnitude of the \textit{conditional average treatment effect} (CATE). Although these two approaches generally lead to similar conclusions regarding decision-making \citep{cdt},  there is currently an illuminating debate on their contrasts, limitations, and bioethical considerations for personalized medicine. We refer the reader to \citet{Sarvet2023,sarvet2024rejoinder} and \citet{MuellerPearl2023,mueller2023perspective} for this debate. 

\paragraph{Main contribution:} We offer contributions to the problems of  identification, partial identification, and doubly-robust, stable inference for the proposed \textit{probability of tiered benefit}. This novel parameter generalizes the PNS to cases where the outcome is either ordered categorical or continuous with a fixed partition of its support. Unlike recent generalizations by \citet{spotify}, \citet{retro}, and \citet{kawakami2024}, our focus is not restricted to queries where strong monotonicity ensures point identification. In fact, we show that this constraint does not suffice for identification when $K \geq 3$. We derive sharp bounds and introduce methods for estimation and inference for these bounds. To address uncertainty quantification issues induced by potential nonregularity, we adapt the \textit{stabilized one-step correction} (S1S) approach by \citet{Luedtke2018} with two extensions. First, our target estimand is a bivariate array containing the bounds, so we employ stabilizing $2\times 2$ matrices rather than scalar weights, and leverage a martingale structure that yields a limiting bivariate Gaussian distribution for a linear transformation of the estimator's asymptotic bias. Second, we evaluate additive corrections using the next out-of-sample unit within each stratum, ensuring valid stratum-specific inference while employing pooled data to estimate population-level nuisance parameters efficiently.

\paragraph{Related literature:} The \textit{probabilities of causation} were formalized in counterfactual terms by \citet{probcausation} for scenarios with binary exposure and outcome. It was shown that point identification can be achieved under a monotonicity assumption. Without this functional constraint, sharp bounds are derived using a combination of observational and experimental data \citep{TP}. Covariate information has been leveraged to define stratum-specific queries and to narrow the bounds through a marginalization step \citep{MuellerLiPearl, li2022unit}. Partial identification has been extended to generalizations of the probabilities of causation in scenarios with nonbinary categorical variables \citep{Li2024, li2024unit}. More recently, a further generalization was proposed for continuous and vector-valued variables, where identification remains achievable under monotonicity constraints \citep{kawakami2024}. Bounds have also been established for the limiting case of the distribution of \textit{individualized treatment effects} (ITE) \citep{Soo}, which is not $g$-identifiable due to the \textit{fundamental problem of causal inference}, as well as for some of its functionals \citep{Firpo,Rusell}.

Robust inference for set-valued or partially identified parameters is an active area of research in statistics and econometrics. Given that bounds are often determined by extrema functionals of distributions, the estimation and inference tasks often target transformations of value functions from individualized optimization problems. Complete conditions for \textit{pathwise differentiability} of such functionals have been characterized, enabling the development of \textit{regular asymptotically linear} (RAL) estimators in a broad class of distributions referred to as \textit{nonexceptional laws} \citep{robins2004optimal,nonuniqueLuedtke}. It has been shown that ambiguities and nonunique solutions in such optimization problems can result in lack of pathwise differentiability, making it impossible to obtain regular and locally unbiased estimators \citep{impossibility}.  Nondifferentiability also poses challenges for other inference frameworks, causing various bootstrap-based methods to become inconsistent \citep{dumbgen1993,fang2019inference}, and making Bayesian credible intervals fail to be asymptotically equivalent to confidence intervals \citep{Kitagawa}.

Various methodological approaches have been proposed to conduct robust semiparametric inference in putative nonregular settings, including: \textit{(i)} targeting thresholded or smoothed surrogates \citep{chakraborty2010}, \textit{(ii)} employing data-adaptive or cross-validated surrogates \citep{bibaut2017data,vanderLaanChapter}, and \textit{(iii)} applying stabilizing weights to sequences of one-step corrections \citep{nonuniqueLuedtke,Luedtke2018}. Other strategies stem from alternative statistical optimality criteria, such as follows: \textit{(iv)} median-bias correction for {intersection bounds} \citep{intbounds,possebom2022}, and \textit{(v)} solutions using {asymptotic minimax regret} \citep{hirano2009,song2014local,Ponomarev,dAdamo}. Recently, the general problem has also been framed in terms of robust inference for conservative bounds \citep{Conservative} and Kantorovich dual bounds \citep{ji2023} from the optimal transport literature. In this context, nonregularities could be potentially addressed via post-hoc feasibility procedures.

\section{The probability of tiered benefit and its identification}\label{sec:problem}


Let $X\in\mathcal{X}$ represent a categorical pre-exposure variable indicating different strata within a population, with $\min_{x\in\mathcal{X}}\pr(X=x)>0$. 

Let $A\in\{0,1\}$ denote a binary point exposure, where $A=1$ signifies \textit{treated} and $A=0$ signifies \textit{not treated}, and $\min_{a\in\{0,1\}}\pr(A=a\mid X=x)>0$ for all $x\in\mathcal{X}$. 

Let $Y$ be the outcome variable and $Y^a$ denote the \textit{potential outcome} under the intervention $\doo(A=a)$. This is the value that $Y$ assumes when $A$ is fixed to have value $a\in\{0,1\}$ within a \textit{structural causal model} (SCM) $\mathcal{M}$. The potential outcome $Y^a$ is a random variable that varies based on unit-level characteristics \citep{PearlCausality,BareinboimHierarchy2022}, and adheres to the consistency axiom, meaning that if $A=a$ and $Y=y$, then $Y^{a}=y$ \citep{robinsConsistency}.

\begin{definition}\label[definition]{def1}
Let $\tilde{Y}\in C=\{C_k\}_{k=1}^K$ be a categorical outcome with $K\geq 2$ tiers, and $\preceq$ be a total order on $ C$, with $\succ$ defined accordingly. The \textit{$x$-specific probability of tiered benefit} $\PB(x)$ is the joint probability of attaining an outcome tier under no treatment $\tilde{Y}^0$ and, counterfactually, a higher outcome tier under treatment $\tilde{Y}^1$, for stratum $x\in\mathcal{X}$. That is: 
\begin{equation}\label{eq:PBd}
    \PB(x):=\pr\left(\bigvee_{k=1}^{K-1} [\tilde{Y}^0= C_k\wedge \tilde{Y}^1\succ C_k] \mid X=x\right) \!= \sum_{k=1}^{K-1}\pr(\tilde{Y}^0= C_k,\tilde{Y}^1\succ C_k \mid X=x).
\end{equation}
\end{definition}

\begin{definition}\label[definition]{def2}
Let $Y\in\mathcal{Y}\subseteq\R$ be an absolutely continuous outcome, with higher values being more desirable. Let $c$ be a partition of the outcome support with fixed thresholds: $\inf\mathcal{Y}=c_0<c_1<\cdots<c_{K-1}<c_{K}=\sup\mathcal{Y}$, thereby producing $K\geq2$ tiers given by disjoint intervals $I_k = (c_{k-1},c_k]$, $k\in[K]:=\{1,\cdots,K\}$. The \textit{$x$-specific probability of tiered benefit} $\PBc(x)$ is the joint probability of attaining an outcome value under no treatment $Y^0$ at any given tier and, counterfactually, an outcome value under treatment $Y^1$ at any higher tier, for stratum $x\in\mathcal{X}$. That is: 
\begin{equation}\label{eq:PBc}
    \PBc(x):=\pr\left(\bigvee_{k=1}^{K-1} [Y^0\in I_k\wedge Y^1>c_k] \mid X=x\right)\! = \sum_{k=1}^{K-1}\pr(Y^0\in I_k,Y^1>c_k \mid X=x).
\end{equation}
\end{definition}

The representation as sum of joint probabilities arises from the mutual exclusion of the counterfactual events involved. Without loss of generality, we will center on the given formulation within the context of a continuous outcome (\cref{def2}). 


The probability of tiered benefit can be computed from a fully specified SCM following a three-step procedure termed \textit{abduction-action-prediction} \citep{PearlCausality,AAP}, or via an augmented graphical model known as a \textit{twin network} \citep{Balke1994b}. Such computations are sensitive to the specification of functional relationships in the SCM. Thus, in cases of \textit{epistemic uncertainty} regarding the causal mechanisms, $\PBc(\cdot)$ cannot be directly computed from the other components of the SCM \citep{probcausation}. Consequently, the probability of tiered benefit is not $g$-identifiable from any unconstrained combination of inputs involving the causal graph, observational data and subsidiary experiments \citep{Robins89,Oberst2019}.

When $K=2$, a single threshold $c_1$ separates the tiers. In this context, $\PBc(x)$ returns the $x$-specific PNS \citep{probcausation,MuellerLiPearl}, given by $\pr(\tilde{Y}^0=0,\tilde{Y}^1=1\mid X=x)=\E[(1-\tilde{Y}^0)\tilde{Y}^1\mid X=x]$, with $\tilde{Y}^a:=\I(Y^a>c_1)$. Although it is not $g$-identifiable, it can be determined under a functional constraint known as \textit{strong monotonicity}.

\begin{assumption}[Strong monotonicity]\label{ass1}
    The potential outcome $Y^a$ is strongly monotonic at stratum $X\!=\!x$, i.e. $\pr(Y^1\!-\!Y^0\geq 0 \mid X\!=\!x)\in\{0,1\}$.
\end{assumption}

This assumption implies that treatment either universally benefits or universally harms individuals within the stratum. This is sufficient for identification of $\operatorname{PNS}(x)$, yielding the value $\operatorname{PNS}(x)=$ $\Delta_a\pr(Y^a>c_1\mid X=x)$ for a nonharmful exposure \citep{TP}, where $\Delta_a$ is the difference operator relative to the binary term $a$. 

For the general $K\geq 2$ case with a nonharmful exposure, strong monotonicity ensures that $\pr(Y^0 > c_k, Y^1\in I_k\mid X\!=\!x) = 0,\ \forall k\in[K-1]$ \citep{spotify}. In other words, the \textit{$x$-specific probability of tiered harm} is zero  \citep{muellerMonotonicity}, where this is defined analogously as $\PHc(x):=\sum_{k=1}^{K-1}\pr(Y^0>c_k,Y^1\in I_k \mid X\!=\!x)$.


\begin{proposition}\label[proposition]{prop:ident}
For $K\geq 3$, strong monotonicity is not sufficient for point identification of the probability of tiered benefit.
\end{proposition}

Justification of \cref{prop:ident} lies in the underdetermined nature of the linear system of counterfactual probabilities $\pr(Y^0\in I_a,Y^1\in I_b\mid X=x)$, where $a,b\in[K]$. This system involves $N_K:=(K-1)K/2$ null entries induced by strong monotonicity; $M_K:=K(K+1)/2$ unknowns; and $L_K := 2K-1$ constraints. The latter accounts for the $K-1$ linearly independent marginal interventional probabilities for each arm, as well as the joint constraint that all cells sum to one. When $K=2$, one has that $M_2=L_2=3$, which allows for a unique solution for the system and hence for $\PBc(x)=\operatorname{PNS}(x)$. Yet, for $K\geq 3$, one has that $M_K>L_K$, which prevents a unique solution. 

One consequence is that, when $K\geq 3$, additional functional constraints on the causal mechanisms beyond monotonicity are necessary to identify $\PBc(x)$ \citep{Cinelli}. Yet, such constraints are hard to derive from domain expertise, and incorrect assumptions can lead to counterintuitive results \citep{Oberst2019,spotify}. As an alternative, we provide bounds $\Lambda(x) \leq \PBc(x) \leq \Upsilon(x)$ both with and without the strong monotonicity constraint. We then propose a semiparametric estimation strategy for these bounds, enabling doubly-robust inference and valid asymptotic uncertainty quantification at both nonexceptional and potentially exceptional laws.

\section{Partial identification}\label{sec:bounds}


Let $W\in\mathcal{W}$ be a vector of pre-exposure variables such that the following two assumptions hold:

\begin{assumption}[Conditional ignorability]\label{ass2}
    $Y^a\indep A\mid W,X,\ \forall a\in\{0,1\}$. 
\end{assumption}

In the SCM framework, this  assumption is implied by \textit{backdoor admissibility} of $\{W, X\}$. That is, if $\mathcal{G}$ is a causal graph on the system's endogenous variables $\mathcal{V}$ containing $\{W,X,A,Y\}$, and $\mathcal{G}[\underbar{A}]$ is the mutilated graph removing the arrows coming out of $A$, then the $d$-separation statement $Y\indep_d A\mid W,X$ in $\mathcal{G}[\underbar{A}]$ implies conditional ignorability \citep{PearlCausality}.

\begin{assumption}[Propensity score positivity]\label{ass3}
    Let $\pi(w,x):=\pr(A=1\mid W=w,X=x)$ be the propensity score, then $ \pi(w,x)\in (0,1)$, $\forall (w,x)\in\mathcal{W}\times\mathcal{X}$ with $p_W(w\mid X=x)\,\pr(X=x) >0$. 
\end{assumption}

Let $S_{k}(w,x,a)$ and $R_{k}(w,x,a)$ be, respectively, the conditional probability of surpassing threshold $c_k$ and the conditional probability of being in tier $I_k$ under treatment arm $A=a$ given $(W=w,X=x)$. That is:
\begin{align}
S_{k}(w,x,a) &:=  \pr(Y>c_k\mid W=w, X=x, A=a),\\
R_{k}(w,x,a) &:=  \pr(Y\in I_k\mid W=w, X=x, A=a) = S_{k-1}(w,x,a) - S_{k}(w,x,a).
\end{align}
 
\begin{proposition}\label[proposition]{prop:monobounds}
For $K\geq 3$, and under strong monotonicity (\cref{ass1}) with a nonharmful treatment, conditional ignorability (\cref{ass2}), and propensity score positivity (\cref{ass3}), sharp bounds for $\PBc(x)$, with $x\in\mathcal{X}$, are given by:
\begin{align}\label{eq:min0}
  \PBc(x) &\geq  \E_{W\mid X=x}\left[S_1(W,x,1) \!-\! R_{K}(W,x,0) \!-\!\sum_{k=2}^{K-1}\min\{R_k(W,x,0);\, R_k(W,x,1)\} \right],\\ \label{eq:max0}
  \PBc(x) &\leq  \E_{W\mid X=x}\left[S_1(W,x,1) \!-\! R_{K}(W,x,0) \!-\!\sum_{k=2}^{K-1}\max\{0;\, R_k(W,x,0)\!+\!R_k(W,x,1)\!-\!1\} \right]
\end{align}
\end{proposition}

\vspace{-0.5cm}
A detailed derivation is provided in \cref{app:1}. 

In certain domains of application, strong monotonicity is an unrealistic assumption, as unintended effects from treatment may occur for some groups of units with positive probability. Therefore, we also provide a set of bounds not relying on strong monotonicity.

\begin{proposition}\label[proposition]{prop:bounds}
For $K\geq 2$, and under conditional ignorability (\cref{ass2}), and propensity score positivity (\cref{ass3}), sharp bounds for $\PBc(x)$, with $x\in\mathcal{X}$, are given by:
\begin{align}\label{eq:min}
  \PBc(x) &\geq  \Lambda(x) := \sum_{k=1}^{K-1}\E_{W\mid X=x}\max\left\{ 0;\,  R_{k}(W,x,0) + S_{k}(W,x,1) - 1\right\},\\ \label{eq:max}
  \PBc(x) &\leq  \Upsilon(x) := \sum_{k=1}^{K-1}\E_{W\mid X=x}\min\left\{ R_{k}(W,x,0);\, S_{k}(W,x,1)\right\}.
\end{align}
\end{proposition}

 A detailed derivation is provided in \cref{app:2}. 

We now focus on estimation and inference for the stratum-specific bounds given by \cref{prop:bounds}. Thus, the target estimand is a bivariate array containing these bounds:
\begin{equation}
    \Psi[P^*](x) := (\Lambda(x);\, \Upsilon(x))^\top\quad \text{for } x\in\mathcal{X}.
\end{equation}

Here, $\Psi:\mathfrak{P}\times\mathcal{X}\rightarrow [0,1]^2$ is a functional that takes a distribution $P$ in a semiparametric model $\mathfrak{P}$, along with a stratum $x\in\mathcal{X}$, and returns the value of the bounds for $\PBc(x)$. We denote with $P^*$ the true joint distribution of the data $O=(W,X,A,Y)$. 


\section{Estimation and inference at a nonexceptional law}\label{sec:inference1}

Certain distributions known as \textit{exceptional laws} may carry ambiguities that lead to nonregularity in the inference problem \citep{robins2004optimal}. Specifically, if for any $k \in [K-1]$, the two arguments of either the $\max$ or $\min$ operators in \cref{eq:min,eq:max} are equal under $P^*$ with positive probability, $\Psi$ becomes nondifferentiable at $P^*$. In such cases, the limiting distribution of the plug-in estimator is nonstandard, and no regular and locally unbiased estimator for $\Psi[P^*]$ exists \citep{impossibility}. Consequently, standard methods ---such as asymptotic, Bayesian, and most bootstrap-based approaches--- fail to yield valid uncertainty quantification  \citep{dumbgen1993,fang2019inference,Kitagawa}.

\begin{proposition}\label[proposition]{prop:diff}
    Let $x\in\mathcal{X}$ and $\mathcal{B}(x)$ be the set containing all $w\in\mathcal{W}$ such that for at least one $k\in[K-1]$ one has that either $R_k(w,x,0)+S_k(w,x,1)=1$ or $R_k(w,x,0)-S_k(w,x,1)=0$. Then, $P^*$ is nonexceptional and $\Psi[\cdot](x)$ is pathwise differentiable at $P^*$ if and only if $\mathcal{B}(x)$ has measure zero, i.e. $\int\I[w\in\mathcal{B}(x)]\,\dd P^*_W(w\mid X=x)=0$.
\end{proposition}

This proposition follows from the extended definitions of exceptional laws and pathwise differentiability by \citet{nonuniqueLuedtke}. 

At a nonexceptional law, estimation and inference can be conducted via a plug-in procedure $\widehat{\Psi}_{\text{plug}}(x) := \Psi[\widehat{P}](x)$, where $\widehat{P}$ couples the empirical distribution of $W \mid X = x$ with an estimated semiparametric model for the conditional outcome distribution. For instance, the latter can be specified as a Bayesian model with Gaussian likelihood, conditional mean $\mu(W,X,A)$ and homoskedastic variance $\sigma^2$. This approach requires the correct specification of such a model for consistency and valid uncertainty quantification. Alternatively, we propose a semiparametric \textit{doubly-robust} method that leverages the propensity score model as well. Under certain technical conditions, this approach provides a consistent estimator even if either the outcome model or the propensity score model is misspecified, as long as both are not incorrect simultaneously \citep{DRbookVariance}.

Let $\{O_i\}_{i\in J}$ be a held-out dataset from the same population, then a doubly-robust estimator of the bounds can be constructed using \textit{one-step corrections} (1S) of the plug-in estimator as follows:
\begin{align}\label{eq:corr1}
        \widehat{\Lambda}_{1S}(x)  &:= \widehat{\Lambda}_{\text{plug}}(x) + \frac{1}{\abs{J(x)}}\sum_{i\in J(x)}\sum_{k=1}^{K-1} \left(\widehat{D}^R_{k}(O_i) + \widehat{D}^S_{k}(O_i)\right)\cdot \lambda_k[\widehat{P}](O_i),\\ \label{eq:corr2}
     \widehat{\Upsilon}_{1S}(x)  &:= \widehat{\Upsilon}_{\text{plug}}(x) + \frac{1}{\abs{J(x)}}\sum_{i\in J(x)}\sum_{k=1}^{K-1} \left[\widehat{D}^R_{k}(O_i)-\left(\widehat{D}^R_{k}(O_i) - \widehat{D}^S_{k}(O_i)\right)\cdot \upsilon_k[\widehat{P}](O_i)\right],
\end{align}

\noindent where $J(x)$ denotes the subset of units within $J$ for which $X=x$, and $\widehat{D}^R_{k},\widehat{D}^S_{k}$ represent the estimated components of the (uncentered) \textit{efficient influence functions} (EIF) for the expectations of $R_k$ under no treatment and of $S_k$ under treatment, respectively, and given by: 
\begin{align}
    \widehat{D}^R_{k}(O_i)  &:= \frac{1-A_i}{1-\widehat{\pi}(W_i,x)}\cdot \left(\mathbb{I}[Y_i\in I_k]-\widehat{R}_{k}(W_i,x,A_i)\right),\\
    \widehat{D}^S_{k}(O_i)  &:= \frac{A_i}{\widehat{\pi}(W_i,x)}\cdot \left(\mathbb{I}[Y_i>c_k]-\widehat{S}_{k}(W_i,x,A_i)\right).
\end{align}

Moreover, $\lambda_k,\upsilon_k$ are \textit{individualized rules} $\lambda_k,\upsilon_k:\mathfrak{P}\times\mathcal{W}\times\mathcal{X}\rightarrow\{0,1\}$ given by:
\begin{align}
\lambda_k[P](w,x) &:= \mathbb{I}[R_{k}(w,x,0) + S_{k}(w,x,1) - 1 > 0],\\
\upsilon_k[P](w,x) &:= \mathbb{I}[R_{k}(w,x,0) - S_{k}(w,x,1)  > 0 ].
\end{align}

A detailed derivation is available in \cref{app:3}. 

Intuitively, these rules pick the solution index of the individual-level optimization problems. For instance,  $\lambda_k[P](w,x)=1$ indicates that the second term in $\max\left\{ 0;\,  R_{k}(w,x,0) + S_{k}(w,x,1) - 1\right\}$ is the maximum of the two inputs. They also determine the "derivatives" of the $\min$ and $\max$ operators. These rules are well-defined and nonambiguous, even when the two terms being compared equate. 




The proposed 1S procedure is analogous to \textit{double/debiased machine learning} (DML) \citep{DML}. Alternatively, at a nonexceptional law, an asymptotically equivalent substitution estimator can also be constructed using \textit{targeted minimum-loss estimation} (TMLE) \citep{TMLEbook1, TMLEbook2}. Confidence and uncertainty regions can be constructed using the EIF-based estimator for the asymptotic covariance of $\widehat{\Psi}_{1S}(x)$ \citep{DRbookVariance}, given by:
\begin{equation}
    \widehat{\Omega}_{1S}(x) := \frac{1}{\abs{J(x)}}\text{Côv}\left(\{\widehat{\Lambda}_{i}+\partial\widehat{\Lambda}_{i};\  \widehat{\Upsilon}_{i}+\partial\widehat{\Upsilon}_{i}\}_{i\in J(x)}\right),
\end{equation}

\noindent where $\widehat{\Lambda}_{i}$ and $\partial\widehat{\Lambda}_{i}$ denote the unit-level prediction from the plug-in estimator and its respective correction for the lower bound; with $\widehat{\Upsilon}_{i}$ and $\partial\widehat{\Upsilon}_{i}$ serving analogous roles for the upper bound.


\section{Estimation and inference at a potentially exceptional law}\label{sec:inference2}

\paragraph{Example 1:} Consider the case of an exposure having zero conditional risk difference for surpassing the first threshold $c_1$  for a subpopulation $(w', x') \in \mathcal{W} \times \mathcal{X}$ with positive probability. In this scenario, the probability of being in the first tier is identical for both treated and untreated individuals in such a group, so $R_1(w', x', 0) = R_1(w', x', 1)$. Since $S_1(w', x', 1) - 1 = -R_1(w', x', 1)$, this  produces a nonregular behavior in the first sum component $\max\{0; R_1(w', x', 0) - R_1(w', x', 1)\}$ of the lower bound.

\paragraph{Example 2:} Suppose the exposure results in a zero conditional risk difference for surpassing the last inner threshold $c_{K-1}$, and that the probability of attaining either of the final two tiers, \(I_{K-1}\) and \(I_{K}\), is the same under no treatment for some subpopulation $(w', x')$ with positive probability. This leads to $R_{K-1}(w', x', 0) = R_{K}(w', x', 0) = S_{K-1}(w', x', 0)$, which creates ambiguity in the last sum term $\min\{S_{K-1}(w', x', 0); S_{K-1}(w', x', 1)\}$ and hence nonregularity in the upper bound.\\

The first example presents a more plausible concern, particularly when prior knowledge indicates that the treatment effect could be zero or minimal, as it only necessitates a null conditional risk difference for surpassing the first inner threshold. In this case, the lower bound is impacted, hindering conservative or pessimistic inference \citep{Conservative}. However, nonregularities may also emerge from unforeseen factors, for instance, when subpopulations have varying immunity to the exposure or when tiers are designed to meet specific balance or fairness criteria.

As demonstrated by \citet{nonuniqueLuedtke, Luedtke2018}, parametric-rate inference and valid uncertainty quantification can still be achieved under certain technical conditions, even at potentially exceptional laws. Here, we adapt and extend the \textit{stabilized one-step correction} (S1S) procedure with two key modifications. While the original approach is designed for population-level queries on the real line and utilizes scalar stabilizing weights, we generalize it to handle bivariate, stratum-specific queries, replacing scalar weights with \textit{stabilizing matrices}. The adapted procedure is summarized in the following steps and detailed in \cref{alg:cap}.

\begin{enumerate}[itemsep=0.5ex,partopsep=0.5ex,parsep=0.5ex]
    \item Permute the units in the dataset. Select an initial data batch size $l>0$.
    \item Learn nuisance parameters $\widehat{P}$ from the initial batch. Identify the stratum for the next unit $x':=X_{l+1}$. Compute the plug-in values $\widehat{\Lambda}(x')$ and $\widehat{\Upsilon}(x')$.
    \item For the next unit and for each unit in the batch within the same stratum $x'$, compute their one-step corrections as given by \cref{eq:corr1,eq:corr2}. 
    \item Calculate $T$: the inverse-root-square of the covariance matrix of the unit-level corrected predictions within the stratum $x'$.
    \item Correct the plug-in estimates of the bounds for stratum $x'$ using the next unit's individual correction. Save it with $T$.
    \item Add the next unit to the batch and repeat from (\textrm{ii}) until reaching the penultimate unit.
    \item The final estimate is the matrix-weighted average of all corrected estimates, weighted by their corresponding stabilizing matrices $T$.
\end{enumerate}

\begin{algorithm}[h]
 \small
\caption{Estimation and inference for $\PBc(\cdot)$ via S1S with stabilizing matrices}\label{alg:cap}
\begin{algorithmic}[1]
\Require data $\{O_i\}_{i=1}^n$, outcome thresholds $c=\{c_k\}_{k=1}^{K-1}$, initial batch size $0<l<n$,\\ set of learners (and super-learning schemes)
\Ensure estimates $\widehat{\Psi}_{S1S}(\cdot)$, and asymptotic covariance matrix $\widehat{\Omega}_{S1S}(\cdot)$
\State Permute data indices; and initialize $n(x)\gets 0$,  $m(x)\gets (0,0)$, and $M(x)\gets 0\cdot \operatorname{Id}_2$,  $\forall x\in\mathcal{X}$
\State \textbf{for} {$j \in \{l,l+1,\dots,n-1\}$}
    \State \phantom{XL} Let $\mathcal{D}\gets [j]$; and learn nuisance parameters of $\widehat{P}$ (cond. outcome and propensity score) from $\mathcal{D}$ 
    \State \phantom{XL} Let $x' \gets X_{j+1}$, $\widehat{\Lambda}(x')\gets \Lambda[\widehat{P}](x')$ and $\widehat{\Upsilon}(x')\gets \Upsilon[\widehat{P}](x')$
    \State \phantom{XXLL} $\widehat{\Lambda}_i\gets \sum_{k=1}^{K-1}\max\left\{ 0;\,  \widehat{R}_{k}(W_i,x,0) + \widehat{S}_{k}(W_i,x,1) - 1\right\}$, \phantom{xxxx} $\forall i\in[j]: X_i=x'$
    \State \phantom{XXLL} $\widehat{\Upsilon}_i\gets  \sum_{k=1}^{K-1}\min\left\{ \widehat{R}_{k}(W_i,x,0);\ \widehat{S}_{k}(W_i,x,1)\right\}$, \phantom{xxxxxxxxxxx} $\forall i\in[j]: X_i=x'$
    \State \phantom{XXLL} $\partial\widehat{\Lambda}_i\gets \sum_{k=1}^{K-1} \left(\widehat{D}^R_{k}(O_i) + \widehat{D}^S_{k}(O_i)\right)\cdot \lambda_k[\widehat{P};O_i]$,  \phantom{xxxxxxxxxxx} $\forall i\in[j+1]: X_i=x'$
    \State \phantom{XXLL} $\partial\widehat{\Upsilon}_i\gets \sum_{k=1}^{K-1} \left[\widehat{D}^R_{k}(O_i)-\left(\widehat{D}^R_{k}(O_i) - \widehat{D}^S_{k}(O_i)\right)\cdot \upsilon_k[\widehat{P};O_i]\right]$,  $\forall i\in[j+1]: X_i=x'$
    \State \phantom{XXLL} $T(x')\gets \text{Côv}\left(\{\widehat{\Lambda}_{i}+\partial\widehat{\Lambda}_{i};\  \widehat{\Upsilon}_{i}+\partial\widehat{\Upsilon}_{i}\}_{i\in \mathcal{D} : X_i=x'}\right) ^{-\frac{1}{2}}$
    \State \phantom{XXLL} $m(x')\gets m(x')+T(x')\left[ (\widehat{\Lambda}(x');\ \widehat{\Upsilon}(x') )^\top+ (\partial\widehat{\Lambda}_{j+1};\ \partial\widehat{\Upsilon}_{j+1})^\top \right]$
    \State \phantom{XXLL} $M(x')\gets M(x')+T(x')$ and $n(x')\gets n(x')+1$
 \Statex \textbf{end for}   
\State $\widehat{\Psi}_{S1S}(x)\gets M(x)^{-1}m(x)$ and   $\widehat{\Omega}_{S1S}(x)\gets n(x)M^{-2}(x)$,  $\forall x\in\mathcal{X}$\\
\Return $\widehat{\Psi}_s(\cdot)$ and  $\widehat{\Omega}_{S1S}(\cdot)$ 
\end{algorithmic}
\end{algorithm}

A detailed derivation with conditions for asymptotic convergence is available in \cref{app:4}.

This procedure requires a sample size big enough to guarantee $n(x)\geq 2$ for all strata, after discounting the initial batch. Uncertainty regions can then be constructed using Monte Carlo sampling from the asymptotic distribution. Let $\{(s_\Lambda, s_\Upsilon)\}_{h=1}^H$ represent samples from a bivariate Gaussian distribution with zero mean and covariance matrix $\widehat{\Omega}_{S1S}(x)$. Define $\widehat{s}(x)$ as the 97.5\% quantile of $\max \{s_\Lambda, -s_\Upsilon\}$. This yields a potentially conservative uncertainty region for $\PBc(x)$ that incorporates both aleatoric and systemic uncertainties, as follows:
\begin{equation}
   \widehat{\Lambda}_{S1S}(x) - \widehat{s}(x) \leq \PBc(x)\leq \widehat{\Upsilon}_{S1S}(x) + \widehat{s}(x),
\end{equation}

\noindent giving the following statistical guarantee:
\begin{equation}
   \liminf_{n\rightarrow\infty}\pr\left(\widehat{\Lambda}_{S1S}(x) - \widehat{s}(x) \leq \PBc(x)\leq \widehat{\Upsilon}_{S1S}(x) + \widehat{s}(x)\right) \geq 0.95.
\end{equation}

\section{Simulations}\label{sec:simulations}

We perform a simulation study in a simplified setup inspired by example 1 in \cref{sec:inference2}, to evaluate the performance of various methods for conducting inference on the bounds of the probability of tiered benefit. Data are generated according to the following SCM:
\begingroup\makeatletter\def\f@size{10.5}\check@mathfonts
\begin{align*}
   & W_1 \sim \operatorname{Unif}(-1,1),\\
   & W_2 = \mathbb{I}\left[v^2>0.25\right],\quad X = \mathbb{I}[v>0],\quad v \sim \operatorname{Unif}(-1,1),\\
    & A = \mathbb{I}\left[0.5\,(W_1+2X-1)+u_A>0\right],\quad  u_A\sim N(0,1),\\
    & Y = (2A-1) + (W_1+X)\cdot [1+0.5(2A-1)] - A\cdot W_2\cdot(W_1+X+2)+u_Y,\ u_Y\sim N(0,4).
\end{align*}
\endgroup

Here $\operatorname{Unif}(a,b)$ denotes the continuous uniform distribution with support $[a,b]$, and $N(\mu,\sigma^2)$ indicates the Gaussian distribution with mean $\mu$ and variance $\sigma^2$. In this setup, we set $K=3$, with fixed thresholds at $c_1=-1.42$ and $c_2=1.09$.

Since the outcome follows a homoscedastic Gaussian distribution, the conditional risk difference, $\Delta_a \pr(Y > c \mid W_1, W_2, X, A = a)$, is zero whenever the CATE, $\Delta_a \E\left[Y \mid W_1, W_2, X, A = a\right]$, is zero, for any value $c$ in the outcome's support. Consider the subpopulation with $W_2 = 1$, which has positive probability in both strata $X \in\{0,1\}$. For this group, the conditional mean outcome is $\E\left[Y \mid W_1, W_2=1, X, A = a\right] = 0.5\, (W_1 + X) - 1$, which is independent of $a$, implying a null CATE. This causes nonregularity in the sum component of the lower bound for the first threshold, impacting the joint inference of the bounds.  In other words, the joint distribution $P^*$ governing the SCM is exceptional due to the existence of a subpopulation, with positive probability, that is immune to treatment.

We benchmark four types of estimators: \textit{(i)} the naïve plug-in estimator, \textit{(ii)} the one-step corrected (1S) estimator from \cref{sec:inference1}, \textit{(iii)} the stabilized one-step corrected (S1S) estimator detailed in \cref{sec:inference2}, and \textit{(iv)} the one-step corrected estimator with \textit{smooth surrogates}. We use a sample size of $n = 5\,000$ and, for the S1S procedure, we set an initial batch size of $l = 2\,000$. For the surrogates, we replace nondifferentiable terms with a smooth variant, i.e., $\max\{0,x\}$ is approximated by $\operatorname{GELU}(x,h) := x \cdot \Phi(x/h)$, where $\Phi$ denotes the Gaussian CDF and $h > 0$ is a smoothing parameter \citep{gelu}. The outcome and propensity score models for all estimators are fit using \textit{multivariate adaptive regression splines} and logistic regression. The true values of $\PBc(\cdot)$ and its bounds are 0.30 with range 0.16--0.69 for $X = 0$, and 0.37 with range 0.25--0.66 for $X = 1$. Results, averaged over 200 iterations, are presented in \cref{tab:simu}.

The 1S procedure consistently outperforms the naïve plug-in estimator across all metrics, especially in enhancing coverage for the lower bound. However, joint coverage remains below the nominal 95\% in both strata. The performance of the smooth surrogate estimator is highly sensitive to the hyperparameter \( h \): at \( h = 0.05 \), it performs similarly to the 1S procedure, but at \( h = 0.15 \), most performance metrics worsen. The S1S estimator improves joint coverage across both strata, particularly for stratum \( X=0 \), though it is slightly conservative in its coverage for stratum \( X=1 \) and exhibits higher MSE compared to the 1S estimator. However, coverage improvement comes at a significantly higher computational cost, as it requires re-learning all nuisance parameters with each new unit added to the data batch.


\begin{table}[t]
\footnotesize
\captionsetup{font=small}
    \centering
    \caption{Performance of proposed procedures and other methods in simulation setup in relation to marginal coverage of confidence intervals for lower bound (Lo.) and upper bound (Up.) of the probability of tiered benefit, joint coverage (Jo.) of bounds, and MSE results. }
    \label{tab:simu}
\begin{tabular}{lcccccccccc}\cmidrule[1pt]{2-11}
& \multicolumn{5}{c}{$X=0$} & \multicolumn{5}{c}{$X=1$}\\ \cmidrule{2-11}
& \multicolumn{3}{c}{Coverage (\%)} & \multicolumn{2}{c}{MSE $\times 10^3$} & \multicolumn{3}{c}{Coverage (\%)} & \multicolumn{2}{c}{MSE $\times 10^3$}\\ \cmidrule{1-11}
Estim. & Lo. & Up. & Jo. & Lo. & Up. & Lo. & Up. & Jo. & Lo. & Up.\\ \midrule
Plug-in estimator  & 0.5 & 59.0 & 0.0 & 0.925 & 0.392 & 40.5 & 82.0 & 37.0 & 0.305 & 0.472\\
One-step (1S) corrected & 95.0 & 77.5 & 75.0 & 0.257 & 0.839 & 96.0 & 93.0 & 91.5 & 0.356 & 0.331\\
1S with GELU($h=.05$) & 95.5 & 66.5 & 64.0 & 0.279 & 1.028 & 96.5 & 92.5 & 91.0 & 0.357 & 0.334\\
1S with GELU($h=.15$) & 70.5 & 46.0 & 28.0 & 0.784 & 2.029  & 91.0 & 85.5  & 75.5 & 0.497 & 0.702\\
\textbf{S1S with stab. matices}  & \textbf{95.5} & \textbf{95.5} & \textbf{92.5} & 0.607 & 0.654 & \textbf{97.0} & \textbf{98.0} & \textbf{95.5} & 0.589 & 0.522\\ \cmidrule[1pt]{1-11}
\end{tabular}
\end{table}


\section{Application case}\label{sec:application}

\begin{figure}[t]
\centering
\footnotesize
    \begin{tikzpicture}
    \begin{axis}[font=\footnotesize,
    scale=0.6,
    height=9.25cm,
    width=6.5cm,  
      xmax=65.0,
      xmin=-10.0,
      ymin=0,
      ymax=15,
      ytick={1,...,14},
      axis y line*=left,
      axis x line*=bottom,
      yticklabels={ , high test score at G5, , low test score at G5, , no prior treatment, , prior treatment, , birth parity > 0, , birth parity = 0, , boys},
      xlabel={percentage points}]
    \addplot+ [red, solid, style=thick, boxplot prepared={
        lower whisker=0.0, upper whisker=13.3 
        }] coordinates {};
    \addplot+ [blue, solid, style=thick, boxplot prepared={
        lower whisker=0.0, upper whisker=20.3
        }] coordinates {}; 
    
    \addplot+ [red, solid, style=thick, boxplot prepared={
        lower whisker=0.0, upper whisker=45.9 
        }] coordinates {};
    \addplot+ [blue, solid, style=thick, boxplot prepared={
        lower whisker=4.63, upper whisker=57.8
        }] coordinates {}; 

    \addplot+ [red, solid, style=thick, boxplot prepared={
        lower whisker=0.0, upper whisker=26.6
        }] coordinates {};
    \addplot+ [blue, solid, style=thick, boxplot prepared={
        lower whisker=0.0, upper whisker=36.7
        }] coordinates {}; 

    \addplot+ [red, solid, style=thick, boxplot prepared={
        lower whisker=0.0, upper whisker=25.8
        }] coordinates {};
    \addplot+ [blue, solid, style=thick, boxplot prepared={
        lower whisker=4.83, upper whisker=53.6
        }] coordinates {}; 

    \addplot+ [red, solid, style=thick, boxplot prepared={
        lower whisker=0.0, upper whisker=28.3
        }] coordinates {};
    \addplot+ [blue, solid, style=thick, boxplot prepared={
        lower whisker=5.0, upper whisker=39.6
        }] coordinates {}; 

    \addplot+ [red, solid, style=thick, boxplot prepared={
        lower whisker=1.9, upper whisker=23.5
        }] coordinates {};
    \addplot+ [blue, solid, style=thick, boxplot prepared={
        lower whisker=4.39, upper whisker=39.6
        }] coordinates {}; 

    \addplot+ [red, solid, style=thick, boxplot prepared={
        lower whisker=0.0, upper whisker=23.2 
        }] coordinates {};
    \addplot+ [blue, solid, style=thick, boxplot prepared={
        lower whisker=3.31, upper whisker=35.8
        }] coordinates {}; 

    \addplot[color=gray,dashed] coordinates {
        		(0,15)
        		(0,0)
        	};
    \addplot[color=gray,dashed] coordinates {
        		(50,15)
        		(50,0)
        	};
    \end{axis}
\end{tikzpicture}\quad
    \begin{tikzpicture}
    \begin{axis}[font=\footnotesize,
    scale=0.6,
    height=9.25cm,
    width=6.5cm,  
      xmax=65.0,
      xmin=-9.5,
      ymin=0,
      ymax=15,
      ytick={1,...,14},
      axis y line*=right,
      axis x line*=bottom,
      yticklabels={, mom edu. > higher, , mom edu. < secondary, , mom age at birth > 28, , mom age at birth < 28, , no missing test, , missing G5 English test, , girls},
      xlabel={percentage points}]
     \addplot+ [red, solid, style=thick, boxplot prepared={
        lower whisker=0.0, upper whisker=21.3
        }] coordinates {};
    \addplot+ [blue, solid, style=thick, boxplot prepared={
        lower whisker=0.00, upper whisker=31.3 
        }] coordinates {}; 

     \addplot+ [red, solid, style=thick, boxplot prepared={
        lower whisker=0.0, upper whisker=31.4
        }] coordinates {};
    \addplot+ [blue, solid, style=thick, boxplot prepared={
        lower whisker=2.32, upper whisker=45.9
        }] coordinates {}; 
        
     \addplot+ [red, solid, style=thick, boxplot prepared={
        lower whisker=0.0, upper whisker=25.9
        }] coordinates {};
    \addplot+ [blue, solid, style=thick, boxplot prepared={
        lower whisker=0.0, upper whisker=36.5
        }] coordinates {}; 
    
     \addplot+ [red, solid, style=thick, boxplot prepared={
        lower whisker=0.0, upper whisker=28.0
        }] coordinates {};
    \addplot+ [blue, solid, style=thick, boxplot prepared={
        lower whisker=3.0, upper whisker=42.2
        }] coordinates {}; 
    
     \addplot+ [red, solid, style=thick, boxplot prepared={
        lower whisker=0.0, upper whisker=26.9
        }] coordinates {};
    \addplot+ [blue, solid, style=thick, boxplot prepared={
        lower whisker=0.0, upper whisker=38.7
        }] coordinates {}; 
        
     \addplot+ [red, solid, style=thick, boxplot prepared={
        lower whisker=0.0, upper whisker=26.7
        }] coordinates {};
    \addplot+ [blue, solid, style=thick, boxplot prepared={
        lower whisker=4.17, upper whisker=49.9
        }] coordinates {}; 

    \addplot+ [red, solid, style=thick, boxplot prepared={
        lower whisker=2.9, upper whisker=33.2
        }] coordinates {};
    \addplot+ [blue, solid, style=thick, boxplot prepared={
        lower whisker=8.0, upper whisker=55.1
        }] coordinates {}; 

    \addplot[color=gray,dashed] coordinates {
        		(0,15)
        		(0,0)
        	};
    \addplot[color=gray,dashed] coordinates {
        		(50,15)
        		(50,0)
        	};
    \end{axis}
\end{tikzpicture}%
\captionsetup{font=small}
\caption{Uncertainty regions for the probability of tiered benefit (\textcolor{blue}{in blue}) and for the probability of tiered harm (\textcolor{red}{in red}) for different strata of the population of Norwegian schoolchildren diagnosed with ADHD, relative to stimulant medication treatment and the mastery tiers for the numeracy national test at eighth school grade. Vertical broken lines indicate 0\% and 50\%, and \textit{G5} indicates fifth school grade.}
\label{fig2}
\end{figure}
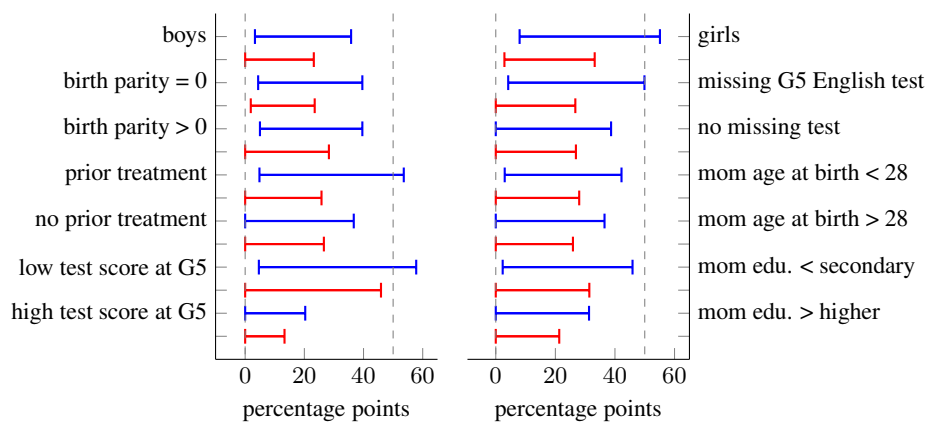

Attention-deficit/hyperactivity disorder (ADHD) is a neurodevelopmental disorder marked by persistent inattention and impulsivity, negatively impacting social, academic, and occupational functioning \citep{icd11}. Affecting approximately 2--6\% of children and adolescents globally, ADHD is among the most common mental health conditions in youth \citep{cortese2023incidence}. Individuals with ADHD often experience negative outcomes, including reduced quality of life and academic underachievement \citep{faraone2021world}. While stimulant medication effectively alleviates symptoms \citep{Cortese2018qd}, its impact on school-related outcomes is modest, with benefits not always translating to improved learning or higher standardized test scores \citep{Storebo2015dz, pelham2022effect}.

In the Norwegian educational system, compulsory tests in numeracy, reading, and English are administered during school grades 5, 8, and 9. Since 2014, test scores have been categorized by the Norwegian Directorate for Education into \textit{mastery tiers}, from tier 1 (poorest) to tier 5 (highest). On average, 12\% of all schoolchildren do not achieve tier 2, and 34\% do not reach tier 3; these figures rise to 26\% and 52\%, respectively, for those with ADHD. Besides, methylphenidate, the most common ADHD medication, can cause adverse side effects such as sleep disturbances and appetite loss \citep{graham2008adverse}, potentially affecting test performance and calling into question the applicability of the monotonicity assumption in this context. 

We apply the developed methods to estimate the probability of tiered benefit and harm from pharmacological treatment for ADHD upon academic achievement, using the official thresholds  among mastery tiers 1, 2, and 3 in the Norwegian numeracy test at eighth grade. As prior evidence suggests a small effect of stimulant medication on academic achievement, applying the stabilized one-step correction (S1S) procedure from \cref{sec:inference2} is well-suited to avoid regularity issues arising from potential null treatment effects in some subpopulations. This analysis uses observational registry data from $9\,352$ individuals and includes a comprehensive set of pre-exposure variables, covering school, family, and child-level information, medical history, sociodemographics, and parental characteristics. Estimated uncertainty regions by the S1S procedure are shown in \cref{fig2}, with additional details on data sources and processing steps provided in \cref{appendixApp}.

These results can be translated into recommendations using the decision-making rules under partial identification proposed by \citet{Cui_2021} and the utility functions defined by \citet{li2022unit}. From a \textit{pessimistic} perspective, the probability of benefit is uniformly small across all strata, peaking at 8\% for girls ---meaning only 8\% of girls would achieve a higher mastery tier with treatment than they would without it---. Meanwhile, the probability of being harmed by stimulant medication can be as high as 33\% for girls, and even higher, reaching 46\% for children who already had low numeracy scores in fifth grade. In contrast, from an \textit{optimistic} viewpoint, the probability of harm is nearly zero across all strata. The groups with the greatest chances of benefiting from medication are children with low numeracy scores in fifth grade (58\%), girls (55\%), and those who received prior ADHD treatment (54\%). From a strict \textit{opportunistic} standpoint, a treatment recommendation cannot be made, as in no group does the minimum benefit exceed the maximum harm. 


The causal interpretation of these results hinges on several graphical and statistical assumptions, some of which are untestable. The potential impacts of violations of some of these assumptions could, in principle, be explored through sensitivity analysis  \citep{diazChapter} and extended partial identification techniques \citep{ding2016sensitivity,vanderweele2017sensitivity}.


\section{Discussion}\label{sec:discussion}

The proposed probability of tiered benefit extends the probability of necessity and sufficiency to new contexts. Though application-agnostic, it can be especially relevant in education and clinical sciences, where outcomes are often tiered and benchmarked against fixed thresholds for purposes like summarization, diagnosis, and intervention evaluation. While partial identification strategies, even when sharp, do not inherently guarantee informative bounds, they provide a more transparent and robust approach compared to imposing strict functional assumptions that may be hard to justify and verify.

The proposed 1S and S1S estimators yield doubly-robust inferential results that outperform the plug-in estimator for bound functionals, even in nonregular scenarios. These approaches could be extended to other bounds in causal inference, including assumption-free bounds for causal effects \citep{probcausation}, instrumental variables \citep{Balke1997}, sensitivity analyses \citep{diazChapter}, and latent confounding \citep{ding2016sensitivity,vanderweele2017sensitivity}.

The S1S procedure provides more reliable results under exceptional distributions, which can emerge from immunities or null treatment effects in certain subpopulations. However, it has some limitations. Firstly, it is computationally intensive. Suppose flexible machine learning models and schemes are implemented, and the running time of the 1S estimator is $\mathcal{O}(n\log n)$ for a large sample size $n$ relative to the number of features. In this case, the running time of S1S estimator, with unit-batch increases, would be $\mathcal{O}(n^2\log n)$, which becomes prohibitively expensive for large $n$. Secondly, because it is not a substitution estimator, it may produce estimates outside the parameter space. To address this, computational efficiency could be improved by integrating and adapting methods from the \textit{online learning} paradigm \citep{van2014online} along with TMLE-based solutions for nondifferentiable parameters \citep{vanderLaanChapter}. We aim to explore these in future work.

\vfill
{\footnotesize
\noindent\textbf{Acknowledgements}: We thank the organizing committee of the Causal Inference Workshop at UAI 2024 in Barcelona for the opportunity to present an early version of this work and receive valuable feedback.

\noindent\textbf{Author contributions}: Johan de Aguas conceptualized the theoretical and methodological research questions, designed and implemented the statistical and algorithmic tools, and drafted the manuscript. Sebastian Krumscheid and Johan Pensar provided feedback on theorems and other mathematical results, and revised the manuscript. Guido Biele proposed the study design, preprocessed the data, provided insights on the application case, and revised the manuscript.

\noindent\textbf{Funding information}: This study was supported by the Research Council of Norway (project number 301081, PI: Guido Biele). Johan Pensar received funding from the Research Council of Norway through Integreat – The Norwegian Centre for Knowledge-Driven Machine Learning (project number 332645). In addition, this work was partially funded by a joint program between the Norwegian Artificial Intelligence Research Consortium (NORA) and the HIDA Helmholtz Information \& Data Science Academy (HIDA) to facilitate a research visit to the Karlsruhe Institute of Technology (KIT).

\noindent\textbf{Ethical approval}: REK ethical approval number 96604.

\noindent\textbf{Conflict of interest}: The authors state no conflict of interest.

\noindent\textbf{Data availability statement}: The data and code utilized for the simulation study are accessible in a personal GitHub repository at \url{https://github.com/johandh2o}. Due to the sensitive nature of the research topic, data from the application case cannot be obtained from the authors, but have to be requested from the relevant Norwegian authorities.
}


\clearpage

\bibliography{bib}


\clearpage


\appendix

\section{Proofs and derivations}

\subsection{Proof of \cref{prop:monobounds}}\label{app:1}

Let $q_{a,b}(w,x):=\pr(Y^0\in I_a,Y^1\in I_b\mid W=w,X=x)$ for $a,b\in[K]$. This is, $q_{a,b}$ is the $(w,x)$-specific joint probability of attaining outcome interval $I_a$ under no treatment and, counterfactually, attaining $I_b$ under treatment. Let $Q(w,x)$ be the matrix containing all these probabilities.

Under strong monotonicity (\cref{ass1}) and a nonharmful treatment, one has that $q_{a,b}=0$ for all $a>b$, and so $Q(w,x)$ can be represented by a lower triangular matrix:
\begin{equation}
  Q(w,x)= 
\begin{tikzpicture}[
  baseline,
  label distance=10pt 
]

\matrix [matrix of math nodes,left delimiter=(,right delimiter=),row sep=0.1cm,column sep=0.1cm, nodes={text width=1.3cm, align=center, minimum height=1.5em}] (m) {
\textcolor{red}{q_{1,1}} &  0    & 0    &  \dots   & 0  & 0   \\ 
\textcolor{blue}{q_{1,2}} &   \textcolor{red}{q_{2,2}}   & 0    &   \dots   & 0 & 0  \\
\vdots & \vdots & \vdots & \ddots & \vdots & \vdots \\
\textcolor{blue}{q_{1,K-2}} & \textcolor{blue}{q_{2,K-2}}     & \textcolor{blue}{q_{3,K-2}} & \dots & 0   & 0   \\
\textcolor{blue}{q_{1,K-1}} & \textcolor{blue}{q_{2,K-1}} &   \textcolor{blue}{q_{3,K-1}}  & \dots & \textcolor{red}{q_{K-1,K-1}}   & 0  \\
\textcolor{blue}{q_{1,K}} & \textcolor{blue}{q_{2,K}}     &   \textcolor{blue}{q_{3,K}}  & \dots & \textcolor{blue}{q_{K-1,K}}   & \textcolor{red}{q_{K,K}}   \\};

\draw[dashed] ($1.05*(m-1-6.north west)$) -- ($1.05*(m-6-6.south west)$);

\draw[dashed] ($0.5*(m-1-1.south west)+0.5*(m-2-1.north west)$) -- ($0.5*(m-1-6.south east)+0.5*(m-2-6.north east)$);


\node[
  fit=(m-6-6)(m-6-6),
  inner xsep=-5pt,inner ysep=3pt,
  below delimiter=\},
  label={[below, xshift=0.0cm, yshift=-1.5cm]$\pr(Y^0\in I_K\mid W=w,X=x)$} 
 ] {};


\node[
  fit=(m-1-6)(m-1-6),
  inner xsep=25pt,inner ysep=2pt,
  right delimiter=\},
  label={[rotate=0, right,  xshift=-1.0cm]$\pr(Y^1\in I_1\mid W=w,X=x)$}
] {};

\end{tikzpicture} 
\end{equation}

All entries in $Q(w,x)$ add up to one and all off-diagonal entries add up to the $(w,x)$-specific probability of tiered benefit $\PBc(w,x)$, thus:
\begin{equation}
    1=\PBc(w,x) + \operatorname{tra} Q(w,x).
\end{equation}

Strong monotonicity implies $q_{1,1}(w,x)=\pr(Y^1\in I_1\mid W=w,X=x)$ and $q_{K,K}(w,x)=\pr(Y^0\in I_K\mid W=w,X=x)$ via fulfillment of the margin constraints. Moreover, under conditional ignorability (\cref{ass2}) and propensity score positivity (\cref{ass3}), these quantities are identified by $R_1(w,x,1)$ and $R_K(w,x,0)$ respectively. Therefore:
\begin{align}
    \PBc(w,x) &= 1 - q_{1,1}(w,x)- q_{K,K}(w,x)-\sum_{k=2}^{K-1}q_{k,k}(w,x)\\
    &= [1-R_1(w,x,1)]  - R_K(w,x,0)-\sum_{k=2}^{K-1}q_{k,k}(w,x)\\
    &= S_1(w,x,1)- R_K(w,x,0) -\sum_{k=2}^{K-1}q_{k,k}(w,x).
\end{align}

The quantities $q_{k,k}(w,x)$ are not identifiable for $k \in \{2, \dots, K-1\}$; however, they can be bounded using the Fréchet inequalities \citep{frechet1951}. 
\begin{align}
   q_{k,k}(w,x) &\geq \max\{0;\ \pr(Y^0\in I_k\mid W=w,X=x)+ \pr(Y^1\in I_k\mid W=w,X=x)-1\},\\
   q_{k,k}(w,x) &\leq \min\{\pr(Y^0\in I_k\mid W=w,X=x);\ \pr(Y^1\in I_k\mid W=w,X=x)\}.
\end{align}

These bounds are sharp, meaning they can be attained and are the narrowest possible in the absence of additional information or constraints. The final bounds for $\PBc(x)$ are given by marginalizing $W$ out after replacing inputs by their identified quantities, resulting in:
\begin{align}
  \PBc(x) &\geq  \E_{W\mid X=x}\left[S_1(W,x,1) \!-\! R_{K}(W,x,0) \!-\!\sum_{k=2}^{K-1}\min\{R_k(W,x,0);\, R_k(W,x,1)\} \right],\\ 
  \PBc(x) &\leq  \E_{W\mid X=x}\left[S_1(W,x,1) \!-\! R_{K}(W,x,0) \!-\!\sum_{k=2}^{K-1}\max\{0;\, R_k(W,x,0)\!+\!R_k(W,x,1)\!-\!1\} \right]
\end{align}

The sharpness of these bounds is preserved through Jensen's inequality, which ensures that beginning with $(w,x)$-specific queries and subsequently applying a marginalization step yields narrower bounds than those obtained by starting with $x$-specific queries, as demonstrated by \citep{MuellerLiPearl}.

\subsection{Proof of \cref{prop:bounds}}\label{app:2}

Let $q_{a,b}(w,x)$ be defined as in \cref{app:1}. By the Fréchet inequalities, one gets:
\begin{align}
   q_{a,b}(w,x) &\geq \max\{0;\ \pr(Y^0\in I_a\mid W=w,X=x)+ \pr(Y^1\in I_b\mid W=w,X=x)-1\},\\
   q_{a,b}(w,x) &\leq \min\{\pr(Y^0\in I_a\mid W=w,X=x);\ \pr(Y^1\in I_b\mid W=w,X=x)\}.
\end{align}

Let $I_a=I_k$ and $I_b=(c_k,c_K)$. Under conditional ignorability (\cref{ass2}) and propensity score positivity (\cref{ass3}), these bounds are identified by:
\begin{align}
  q_{a,b}(w,x) &\geq  \max\left\{ 0;\,  R_{k}(w,x,0) + S_{k}(w,x,1) - 1\right\},\\ 
  q_{a,b}(w,x) &\leq  \min\left\{ R_{k}(w,x,0);\, S_{k}(w,x,1)\right\}.
\end{align}

By marginalizing $W$ out and summing over all $k\in[K-1]$, one gets:
\begin{align}
  \PBc(x) &\geq  \Lambda(x) := \sum_{k=1}^{K-1}\E_{W\mid X=x}\max\left\{ 0;\,  R_{k}(W,x,0) + S_{k}(W,x,1) - 1\right\},\\ 
  \PBc(x) &\leq  \Upsilon(x) := \sum_{k=1}^{K-1}\E_{W\mid X=x}\min\left\{ R_{k}(W,x,0);\, S_{k}(W,x,1)\right\}.
\end{align}

\subsection{Derivation of one-step corrected estimator}\label{app:3}

Consider a parametric submodel $P_\epsilon\in\mathfrak{P}$ indexed by a small fluctuation parameter $\epsilon\in\R$ and a point-mass contamination $O_i=(W_i,X_i,A_i,,Y_i)$, such that $P^\epsilon(O)=\epsilon\,\I({O}={O}_i)+(1-\epsilon)\,P^*({O})$. Under some technical conditions involving \textit{(i)} fully nonparametric or saturated model $\mathfrak{P}$, \textit{(ii)} smoothness of the paths within the model, \textit{(ii)} positivity of the propensity score and of each stratum $x\in\mathcal{X}$, and \textit{(iv)} boundedness of the outcome mean, the Gâteaux derivative, and their variances, we have that $\Lambda[\cdot](x)$ is pathwise differentiable at nonexceptional law $P^*\in\mathfrak{P}$  \citep{hines2022demystifying}. Its \textit{efficient influence function} (EIF) at $P^*$ being evaluated at point $O_i\sim P^*$, with $X_i=x$, can be computed using the chain rule and gradient algebra for the Gâteaux derivative. Thus:
\begin{align}
     \dv{}{\epsilon}\Lambda[P^\epsilon](x) \eval_{\epsilon=0}   & = \sum_{k=1}^{K-1}\dv{}{\epsilon}\E_{W\mid X=x}^\epsilon\max\left\{ 0;\,  R^\epsilon_{k}(W,x,0) + S^\epsilon_{k}(W,x,1) - 1\right\}\eval_{\epsilon=0} \\ \label{eq:app1}
    & = \sum_{k=1}^{K-1}\left(D_k^R(O_i)+D_k^S(O_i)\right) \cdot \lambda_k[P^*](W_i,x)\\ \nonumber
    &\quad +\sum_{k=1}^{K-1}\max\left\{ 0;\,  R^*_{k}(W_i,x,0) + S^*_{k}(W_i,x,1) - 1\right\} -\Lambda[P^*](x).
\end{align}

The expression in line \eqref{eq:app1} is justified by the application of the chain rule at nonexceptional $P^*$, and so the term $\max\left\{ 0;\,  R_{k}(W,x,0) + S_{k}(W,x,1) - 1\right\}$ is differentiable \textit{almost everywhere}, with the functional derivative being the Heaviside step $\lambda_k(W,x):=\mathbb{I}[R_{k}(W,x,0) + S_{k}(W,x,1) - 1>0]$.

Let $D^\Lambda(O_i):=\dv{}{\epsilon}\Lambda[P^\epsilon](x) \eval_{\epsilon=0}$ evaluated at $P^*$ and observation $O_i\sim P^*$ with $X_i=x$. One key property of the EIF is that it satisfies the moment condition $\E_{O\mid X=x} D^\Lambda(O)=0$, making its empirical counterpart suitable as an estimating equation for $\Lambda[P^*](x)$. Let $\widehat{P}$ be an initial estimator of $P^*$ and $J$ an evaluation dataset. Then, a one-step corrected estimator of $\Lambda[P^*](x)$, denoted $\widehat{\Lambda}_{1S}(x)$, can be constructed as a solution to the empirical moment condition of the EIF, namely:
\begin{equation}
\begin{split}
   \frac{1}{\abs{J(x)}}\sum_{i\in J(x)} \widehat{D}^\Lambda (O_i) = \frac{1}{\abs{J(x)}}\sum_{i\in J(x)}\sum_{k=1}^{K-1} \left(\widehat{D}^R_{k}(O_i) + \widehat{D}^S_{k}(O_i)\right)\cdot \lambda_k[\widehat{P}](O_i) \\+ \underbrace{\frac{1}{\abs{J(x)}}\sum_{i\in J(x)}\sum_{k=1}^{K-1}\max\left\{ 0;\,  \widehat{R}_{k}(W_i,x,0) + \widehat{S}_{k}(W_i,x,1) - 1\right\}}_{\widehat{\Lambda}_{\text{plug}}(x)}- \widehat{\Lambda}_{1S}(x) = 0,
\end{split}
\end{equation}

\noindent where $J(x)$ denotes the subset of indices within $J$ for which $X=x$. By rearranging terms, one gets:
\begin{equation}
        \widehat{\Lambda}_{1S}(x)  = \widehat{\Lambda}_{\text{plug}}(x) + \frac{1}{\abs{J(x)}}\sum_{i\in J(x)}\sum_{k=1}^{K-1} \left(\widehat{D}^R_{k}(O_i) + \widehat{D}^S_{k}(O_i)\right)\cdot \lambda_k[\widehat{P}](O_i).
\end{equation}

A one-step corrected estimator formulation for the upper bound  $\widehat{\Upsilon}_{1S}(x)$ follows analogously.

\subsection{Derivation of stabilized one-step corrected estimator}\label{app:4}

Even when the rules $\lambda_k,\upsilon_k$ are well-defined by enforcing a strictly positive condition, ambiguities introduced by exceptional laws may cause their estimates to remain unstable, even as the sample size grows indefinitely. If these rules were instead \textit{known} and not estimated from the data, $\Psi[\cdot]$ would be pathwise differentiable at any $P\in\mathfrak{P}$, given certain technical conditions presented in  \cref{app:3}.

We handle the data-dependent nature of the estimated rules by building upon \textit{stabilized one-step correction} (S1S) approach by \citet{nonuniqueLuedtke, Luedtke2018}. This setup enables us to treat a sequence of rule estimates as \textit{known}, allowing a martingale version of the central limit theorem (CLT) to characterize the limiting distribution of the estimator.

To simplify the notation, assume that all functionals, distributions, and samples henceforth are conditioned on a given stratum $X = x$. Let $\Psi[P;\lambda,\upsilon]:= (\Lambda[P;\lambda] ; \Upsilon[P;\upsilon])^\top$ be the evaluation of $\Psi[P]$ when the rules are fixed at given $\lambda=\{\lambda_k\}_{k}^{K-1}$ and $\upsilon=\{\upsilon_k\}_{k}^{K-1}$, with:
\begin{align}
\Lambda[P;\lambda] &:= \sum_{k=1}^{K-1}\E_{W|X=x}\left[R_{k}(w,x,0) + S_{k}(w,x,1) - 1\right]\cdot \lambda_k(w,x), \\
\Upsilon[P;\upsilon] &:= \sum_{k=1}^{K-1}\E_{W|X=x}\left\{R_{k}(w,x,0) - \left[R_{k}(w,x,0) - S_{k}(w,x,1)\right]\cdot \upsilon_k(w,x)\right\}.
\end{align}

This reparametrization is consistent in the sense that $\Psi[P^*;\lambda^*,\upsilon^*] = \Psi[P^*]$, where the asterisk indicates true values.

Let $P^n$ denote the empirical distribution of $O_{1:n}=\{O_i\}_{i=1}^n$, with each observation being an independent sample from $P^*$, and $\lambda^n,\upsilon^n$ the estimated rules from $P^n$. Let $\partial\Psi^n(O)$ be the bivariate one-step corrections given by \cref{eq:corr1,eq:corr2} at $P^n$ and observation $O$. Then, we can express the bias of the corrected estimate $\widehat{\psi}^n$ as follows:
\begin{align}
    \widehat{\psi}^n-\Psi[P^*] &= \Psi[P^n;\lambda^n,\upsilon^n] + \partial\Psi^n(O_{n+1}) - \Psi[P^*] \\
    &= \Psi[P^*;\lambda^n,\upsilon^n] + \partial\Psi^n(O_{n+1}) - \Psi[P^*] + \operatorname{Rem}(P^n)  - \E^*\left[D^n(O)\mid O_{1:n} \right].
\end{align}

Here, $\operatorname{Rem}(P^n):=\Psi[P^n;\lambda^n,\upsilon^n]-\Psi[P^*;\lambda^n,\upsilon^n] +  \E^*\left[D^n(O)\mid O_{1:n} \right]$ represents the second-order remainder term from the von Mises expansion of the target at $P^n$ with fixed rules. Under an exceptional law, the variance of the estimated bias, minus the second-order remainder, is unstable, as it fluctuates with each new observation due to underlying ambiguities. However, by fixing the rules based on accumulated observations $O_{1:n}$ and deriving uncertainty from the next observation $O_{n+1}$, a bivariate martingale structure is introduced. This setup enables the use of a generalized CLT that applies weights determined by the inverse \textit{matrix standard deviation} ---the inverse square-root covariance matrix--- of each new  realization, $T_n=\text{Côv}( {\Psi}^n(O_n) + \partial\Psi^n(O_{n}) )^{-1/2}$. Hence, for $0<l<n$ and $M_n=\sum_{j=l}^{n-1} T_j$, one has that, as $(n-l)\rightarrow\infty$, $(n-l)^{1/2} M_n^{-1} \sum_{j=l}^{n-1} T_j (\widehat{\psi}^j-\Psi[P^*])$ converges to a bivariate Gaussian distribution with a unit covariance matrix under the following consistency and boundedness conditions \citep{nonuniqueLuedtke, Luedtke2018}:
\begin{enumerate}
    \item $(n-l)^{-\frac{1}{2}}\sum_{j=l}^{n-1}T_j\, {\operatorname{Rem}}({P}^j)\overset{\pr}{\rightarrow} (0,0)$, 
    \item $(n-l)^{-1}\sum_{j=l}^{n-1}\abs{T^2_j\,\text{Cov}\left(\{{\Psi}^j(O_i) + \partial\Psi^j(O_{i})\}_{i=1}^j\right)-\operatorname{Id}_2}\overset{\pr}{\rightarrow} 0\cdot \operatorname{Id}_2$, 
    \item $\exists\, \xi<\infty: (n-l)^{-1}\sum_{j=l}^{n-1}\pr\left(\norm{T\,(\Psi^j(O_{j+1}) + \partial\Psi^j(O_{j+1}))^\top}_2<\xi \mid O_{1:j-1} \right)\overset{\pr}{\rightarrow} 1$. 
\end{enumerate}

The mean of this limiting distribution is influenced by the asymptotic behavior of $\Psi[P^*;\lambda^n,\upsilon^n]-\Psi[P^*]$. For the lower bound, however, this mean is guaranteed to be less than or equal to zero, as it reflects the gap to the true maximum. To construct \textit{potentially conservative} one-sided confidence intervals from below for both bounds, one can either target the negative of the upper bound or invert the sign of samples for the upper bound component from the asymptotic distribution.

In contrast to the 1S estimator discussed in \cref{sec:inference1}, the S1S approach eliminates the need for sample splitting. This is due to the martingale process structure, which inherently depends on \textit{fixed} past data and out-of-sample evaluations \citep{van2014online}.

\section{Data details for application case}\label{appendixApp}

We assess the benefits from pharmacological treatment with stimulant medication upon the numeracy test performance at grade 8 obtained by Norwegian children diagnosed with ADHD. By integrating information from national registries, we compile data on the medication history and national test scores of all children diagnosed with ADHD born between 2000 and 2007 in Norway, who would go to take the national test up to 2021. We exclude those with severe comorbid disorders and those with missing test scores at grade 5 and impute the missing values at grade 8 (totaling $9\,352$ individuals). Variables at the student, family, and school levels are linked from the Norwegian Prescription Database (NorPD), the Norwegian Patient Registry (NPR), the Database for Control and Payment of Health Reimbursement (KUHR), Statistics Norway (SSB), and the Medical Birth Registry of Norway (MBRN). We leverage data on students' and parents' diagnoses and their consultations with medical services during pre-exposure and post-exposure periods. Indicators of post-exposure medical status serve as proxies for adverse effects of treatment and its consequences. To operationalize relevant variables, we employ the following grouping:
\begin{itemize}
    \item \textbf{pre-exposure covariates} $W,X$: sex at birth, birth year/month cohorts, birth parity number, raw scores at grade 5 national test for numeracy and reading, missingness indicator for scores at grade 5 English national test, mother's education level, mother's age at birth, student's and parents' diagnoses and medical consultations for related comorbid disorders, school identification (fixed effect), prior dispensations of ADHD stimulant medication for at least 90 days, and duration of prior treatment.
    \item \textbf{Exposure} $A$: 
    having received dispensations of ADHD stimulant medication for at least 75\% of the prescribed treatment period between the start of grade 6 and the national test in grade 8.
    \item \textbf{Outcomes} $Y$: raw scores at grade 8 national test for numeracy and corresponding official thresholds for tiers 1, 2 and 3.
\end{itemize}

\end{document}